\documentclass[aps, pra, showpacs, twocolumn]{revtex4}

\usepackage[latin1]{inputenc}
\usepackage[T1]{fontenc}
\usepackage[english]{babel}

\usepackage{graphicx}
\usepackage{braket}

\usepackage[intlimits]{empheq}
\usepackage{amssymb}
\usepackage{amsthm}
\usepackage{mathtools}
\usepackage{amsfonts}

\allowdisplaybreaks[1]

\usepackage{hyperref}

\newcommand{\Axy}[2]{\hat{A}_{#1}^{\ #2}}

\newcommand{\Ann}{\hat{A}_{n}^{\ n}}

\newcommand{\aop}[1]{\hat{a}_{#1}}
\newcommand{\aopd}[1]{\hat{a}^{\dag}_{#1}}

\newcommand{\bop}[1]{\hat{b}_{#1}}
\newcommand{\bopd}[1]{\hat{b}^{\dag}_{#1}}

\newcommand{\cop}[1]{\hat{c}_{#1}}
\newcommand{\copd}[1]{\hat{c}^{\dag}_{#1}}

\newcommand{\dop}[1]{\hat{d}_{#1}}
\newcommand{\dopd}[1]{\hat{d}^{\dag}_{#1}}

\newcommand{\eop}[1]{\hat{e}_{#1}}
\newcommand{\eopd}[1]{\hat{e}^{\dag}_{#1}}

\begin{document}

\title{Phase transitions and dark state physics in two color superradiance}
\author{Mathias Hayn}
\author{Clive Emary}
\author{Tobias Brandes}
\affiliation{Institut f\"ur Theoretische Physik, Technische Universit\"at Berlin, 10623 Germany}
\date{\today}

\begin{abstract}
	We theoretically study an extension of the Dicke model, where the single-particle Hamiltonian has three energy levels in Lambda-configuration, i.e. the excited state is coupled to two non-degenerate ground states via two independent quantized light fields. The corresponding many-body Hamiltonian can be diagonalized in the thermodynamic limit with the help of a generalized Holstein--Primakoff transformation. Analyzing the ground-state energy and the excitation energies, we identify one normal and two superradiant phases, separated by phase transitions of both first and second order. A phase with both superradiant states coexisting is not stable. In addition, in the limit of two degenerate ground states a dark state emerges, which seems to be analogous to the dark state appearing in the well known stimulated Raman adiabatic passage scheme.
\end{abstract}


\maketitle

%-----------------------------------------------------

\section{Introduction}

Superradiance is a collective phenomenon originating from atomic physics. There, it is regarded as a collective spontaneous emission process of a dense ensemble of radiating atoms~\cite{Gross1982}. The atoms interact indirectly via a light field. The first microscopic description of this phenomenon was given by Dicke~\cite{Dicke1954}.

In the context of phase transitions a collection of two-level systems coupled linearly to one scalar bosonic mode undergoes a second-order phase transition from a normal to a superradiant phase at a certain critical coupling strength. This phase transition has been investigated theoretically a long time ago by Hepp and Lieb~\cite{Hepp1973} and also by Wang and Hioe~\cite{Wang1973}. However, there is no experimental realization in atomic systems to date. There were theoretical proposals to produce this phase transition in artificial quantum systems like \textit{circuit} or \textit{cavity} \textit{quantum electrodynamic} (QED) systems~\cite{Chen2007, Dimer2007, Lambert2009}. Though, there exist no-go theorems for atomic, cavity and circuit QED systems which theoretically preclude the normal--superradiant phase transition~\cite{Rzazewski1975, Nataf2010, Viehmann2011}.

Recently, experimental progress was achieved in this field by the group of Esslinger, who coupled a Bose--Einstein condensate to a single mode of an open optical cavity~\cite{Baumann2010}. The unitary dynamics of this system is described by an effective Dicke Hamiltonian~\cite{Baumann2010, Nagy2010}. Experimentally, the normal--superradiant phase transition is observed by measuring the mean intracavity photon number.

Inspired by this experimental realization of an effective Dicke-Hamiltonian, in this paper we theoretically investigate an extension of the Dicke model. Here, three-level systems in $\Lambda$-configuration are considered, which are coupled to two independent scalar bosonic modes. We are interested in how the phase transition is changed in this configuration. Furthermore, coherent population trapping~\cite{Arimondo1996}, dark states and the \textit{stimulated Raman adiabatic passage} (STIRAP) scheme~\cite{Bergmann1998} are associated with this kind of system in the single-particle and semi-classical case. We therefore study to what extent dark state physics plays a role in our quantum-many-body setting.

The paper is organized as follows: At the beginning, in Sec.~\ref{sec:Model} we introduce the model, give a detailed description of the Hamiltonian and discuss the symmetries of the model. Subsequently in Sec.~\ref{sec:Methods} we describe the Holstein--Primakoff transformation for multilevel systems and derive an effective Hamiltonian in the thermodynamic limit. We diagonalize this effective Hamiltonian and give explicit expressions for the ground-state energy and the excitation energies. Section~\ref{sec:Discussion} addresses the phase transition: The zero-temperature phase diagram is mapped out and analysed. In Sec~\ref{sec:DarkState} we discuss properties of the appearing dark state. Finally, Sec.~\ref{sec:Conclusions} closes our contribution with some conclusions.

%-----------------------------------------------------

%-----------------------------------------------------

\section{The model}
\label{sec:Model}

\begin{figure}[t]
	\includegraphics[width=8cm]{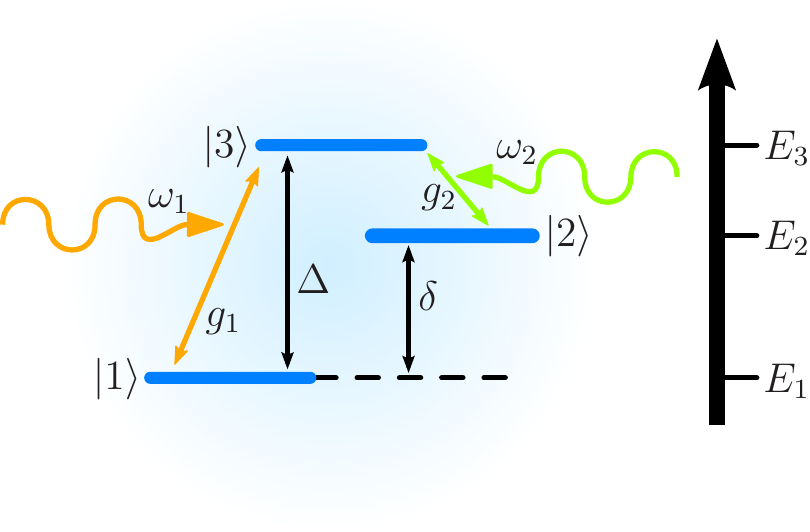}
	\caption{(Color online) Level structure of the $\Lambda$-configuration. One particle has two ground states $\ket{1}$, $\ket{2}$ and one excited state $\ket{3}$, where the excited state is coupled to the two ground states via two independent scalar bosonic modes with in general different frequencies $\omega_1$, $\omega_2$ and coupling strengths $g_1$, $g_2$.}	\label{fig:model}
\end{figure}

We consider a quantum mechanical system consisting of $\mathcal{N}$ distinguishable particles and two independent scalar bosonic modes. Each particle $i$ possess three energy levels $\ket{1}^{(i)}$, $\ket{2}^{(i)}$, and $\ket{3}^{(i)}$ with energies $E_1 \leq E_2 \leq E_3$, respectively. For later analysis we define, $\Delta = E_3 - E_1$, $\delta = E_2 - E_1$, with $\Delta \ge \delta \ge 0$. The level scheme is in so-called \textit{Lambda}($\Lambda$)-configuration. Each of the two lowest energy levels couple to the highest energy level via one of the bosonic modes respectively (see Fig.~\ref{fig:model}). The Hamiltonian has the form ($\hbar=1$)

\begin{multline}	\label{eq:HamOp}
	\hat{H} = \sum_{n=1}^3 E_n \, \Ann + \sum_{n=1}^2 \Bigl\{ \omega_n \, \hat{a}_n^\dag \, \hat{a}_n \\
	+ \frac{g_n}{\sqrt{N}} \bigl( \hat{a}_n^\dag + \hat{a}_n \bigr) \, \Bigl( \Axy{n}{3} + \Axy{3}{n} \Bigr) \Bigr\}.
\end{multline}
Here, $\Axy{r}{s}$ are defined by
\begin{equation}	\label{eq:DefAnm}
	\Axy{r}{s} = \sum_{i=1}^{\mathcal{N}} \ket{r}^{(i)} \negthickspace \bra{s}, \quad r,s = 1, 2, 3,
\end{equation}
and represent collective particle operators.

The diagonal operators $\Ann$ measure the occupation of the $n$th energy level, i.e. how many of the $\mathcal{N}$ particles are in the energy state $\ket{n}$. This illustrates the first term in the Hamiltonian~\eqref{eq:HamOp}. The second term gives the energy of the two scalar bosonic modes, each one having the excitation energy $\omega_1$ and $\omega_2$ respectively. The operators $\aopd{n}$ and $\aop{n}$ create and annihilate a boson in the $n$th mode. They fulfill canonical commutator relations, $[ \aop{n}, \aopd{m}] = \delta_{n,m}$ and $[\aop{n}, \aop{m}] = 0$. Lastly, the third term in the Hamiltonian~\eqref{eq:HamOp} represents the interaction of the particles with the two bosonic modes. Here, $g_n$ are the corresponding coupling constants.

We call the first $\ket{1}$ and the third $\ket{3}$ energy levels of the particle system together with the first bosonic mode \textit{blue branch}, since $\Delta \ge \delta$ is assumed. Correspondingly, we call the second $\ket{2}$ and the third $\ket{3}$ energy level of the particle system together with the second bosonic mode \textit{red branch} (cf. Fig.~\ref{fig:model}).

\textit{Symmetries and phase transition.}
Our Model is a generalization of the Dicke model~\cite{Dicke1954, Emary2003}, where particles with only two energy levels are considered, and the two states are coupled via one scalar bosonic mode. In the thermodynamic limit the Dicke model exhibits a non-analytic behavior in physical observables as a function of the coupling strength $g$. Thus, the Dicke model exhibits a (quantum) phase transition, which is continuous, i.e. of second order and separates two phases: a \textit{normal phase} and a so-called \textit{superradiant phase}. The superradiant phase has a ground state with spontaneously broken symmetry. A similar behavior is anticipated in the extended model.

In analogy to the Dicke model, here exist two symmetry operators

\begin{equation}	\label{eq:DefParity}
	\hat{\Pi}_n = \exp \Bigl\{ -i \pi \bigl( - \Ann + \aopd{n} \, \aop{n} \bigr)  \Bigr\}, \qquad n = 1, 2,
\end{equation}
which commute with the Hamiltonian~\eqref{eq:HamOp}. These operators have the physical meaning of \textit{parity operators} and have eigenvalues $\pm 1$. The operator $\hat{\eta}_n = -\Ann + \aopd{n} \, \aop{n}$ in the exponent of the parity operator~\eqref{eq:DefParity} is related to the number of excitations in the blue ($n=1$) or in the red ($n=2$) branch of the $\Lambda$-system and the number of excitations in the corresponding $n$th bosonic mode, respectively. The operator $\hat{\eta}_n$ itself is not conserved, i.e. $[\hat{\eta}_n, \hat{H}] \neq 0$. This is consistent with the Dicke model~\cite{Emary2003}. In the rotating wave approximation the operators $\hat{\eta}_n$ become conserved quantities. Conservation of the two parities means that the Hilbert space decomposes into four irreducible subspaces. It is the parity which is spontaneously broken in the superradiant phase of the Dicke model. Thus, we expect that at least \textsl{one} of the parities is also spontaneously broken in our model.

Using the definition~\eqref{eq:DefAnm} of the operators $\Axy{r}{s}$, one can show that the two sets of operators $\bigl\{ \frac{1}{2} ( \Axy{3}{3} - \Axy{n}{n}), \Axy{n}{3}, \Axy{3}{n} \bigr\}$, $n = 1, 2$ fulfill the angular momentum algebra respectively, i.e. they are generators of the special unitary group SU(2) and can be understood as angular momentum operators. In addition, the operators $\Axy{r}{s}$ fulfill the algebra of generators of the unitary group U(3)~\cite{Okubo1975, Klein1991}

\begin{equation}
	\bigl[ \Axy{r}{s} , \Axy{n}{m} \bigr] = \delta_{s,n} \, \Axy{r}{m} - \delta_{r,m} \, \Axy{n}{s}
\end{equation}
and are, according to that, generators of the group U(3). It is known, that the generators of the group U(N) can be represented by either N or by $\text{N} - 1$ independent bosons~\cite{Okubo1975, Klein1991}. The first choice corresponds to the \textit{Schwinger boson} representation~\cite{Okubo1975}, the latter choice to the \textit{Holstein--Primakoff} transformation of the generators~\cite{Okubo1975, Holstein1940, Klein1991}.

%-----------------------------------------------------

%-----------------------------------------------------

\section{Methods}
\label{sec:Methods}

The Dicke model was introduced in 1954~\cite{Dicke1954}. To date there exists no exact analytical solution to this model for a finite number $\mathcal{N}$ of particles. However, the Dicke Hamiltonian can be exactly diagonalized in the thermodynamic limit~\cite{Emary2003}, i.e. $\mathcal{N} \rightarrow \infty$. This can be achieved by using the already mentioned Holstein--Primakoff transformation. In this article we apply a generalized version of the Holstein--Primakoff transformation to diagonalize the Hamiltonian~\eqref{eq:HamOp} of the $\Lambda$-system.

\subsection{Generalized Holstein--Primakoff transformation}
In the present paper we discuss the $\Lambda$-system which has $\text{N}=3$ single-particle states. Though, we will formulate the following argument for a general number N of single-particle states. The number of particles is denoted by $\mathcal{N}$, whereas the number of single-particle states is denoted by N.

The generalized Holstein--Primakoff transformation maps the generators $\Axy{r}{s}$ of the group U(N) onto a combination of creation and annihilation operators $\bopd{r}, \bop{r}$ of $\text{N} - 1$ independent bosons. Hence, the operators $\bopd{r}$ and $\bop{r}$ fulfill canonical commutator relations, $[\bop{r}, \bopd{s}] = \delta_{r,s}$, $[\bop{r}, \bop{s}] = 0$. These bosons we will refer to as Holstein--Primakoff bosons (HP bosons).  One of the N states of the one particle system is called the \textsl{reference state}, which we denote with $\ket{m}$. The meaning of the state $\ket{m}$ and which of the N states can be used as a reference state will be elucidated later. Then, the generalized Holstein--Primakoff transformation is given by~\cite{Klein1991}

\begin{equation}	\label{eq:DefHP}
\left.
\begin{aligned}
	\Axy{r}{s} &= \bopd{r} \, \bop{s}, \\
	\Axy{r}{m} &= \bopd{r} \, \hat{\Theta}_m \bigl( \mathcal{N} \bigr), \\
	\Axy{m}{s} &= \hat{\Theta}_m \bigl( \mathcal{N} \bigr) \, \bop{s}, \\
	\Axy{m}{m} &= \hat{\Theta}_m \bigl( \mathcal{N} \bigr)^2,
\end{aligned}
\quad \right\}
\; r,s \neq m
\end{equation}
with
\begin{equation}	\label{eq:DefTheta}
	\hat{\Theta}_m \bigl( \mathcal{N} \bigr) = \sqrt{\mathcal{N} - \sum_{r \neq m} \bopd{r} \, \bop{r} \hphantom{a}}.
\end{equation}
There are at most $\mathcal{N}$ HP bosons per mode, i.e. the expectation value satisfies $\braket{\bopd{r} \, \bop{r}} \le \mathcal{N}$, $r \neq m$, due to the operator $\hat{\Theta}_m \bigl( \mathcal{N} \bigr)$ and the fact that $\bop{r}$ acting on a state with zero HP bosons in the $r$th mode equals to zero. In addition, the number of HP bosons in all $\text{N} - 1$ modes does not exceed $\mathcal{N}$, i.e. $\sum_{r \neq m} \braket{\bopd{r} \, \bop{r}} \le \mathcal{N}$.

\begin{figure}[tb]
	\includegraphics[width=4cm]{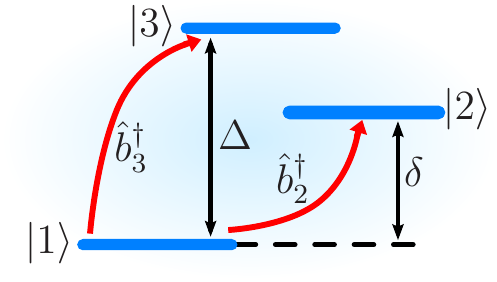}
	\includegraphics[width=4cm]{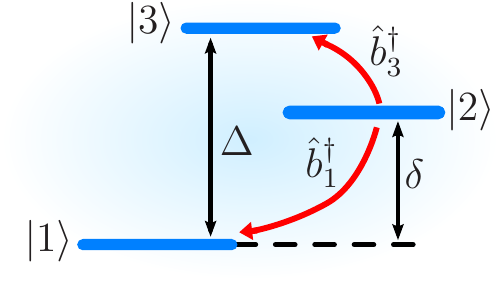}
	\caption{(Color online) Physical interpretation of the bosons introduced via the generalized Holstein--Primakoff transformation~\eqref{eq:DefHP}: The two bosonic operators $\bopd{r}$, with $r \neq m$, can be understood as collectively exciting the particles from the reference state $\ket{m}$ (\textit{left}: $m=1$, \textit{right}: $m=2$) to the state $\ket{r}$. The analogue holds for the annihilation operators $\bop{r}$.\label{fig:HPBosonen}}
\end{figure}

We now apply the generalized Holstein--Primakoff transformation~\eqref{eq:DefHP} to the Hamiltonian~\eqref{eq:HamOp} with e.g. $\ket{1}$ as the reference state ($m=1$) and obtain
\begin{multline}	\label{eq:HamOpHP}
	\hat{H}_{m=1} = E_1 \, \mathcal{N} + \delta \, \bopd{2} \, \bop{2} + \Delta \, \bopd{3} \, \bop{3} + \sum_{n=1}^2 \omega_n \, \aopd{n} \, \aop{n} \\
	+ \frac{g_1}{\sqrt{\mathcal{N} \,}} \bigl( \aopd{1} + \aop{1} \bigr) \, \left( \bopd{3} \, \hat{\Theta}_1 \bigl( \mathcal{N} \bigr) + \hat{\Theta}_1 \bigl( \mathcal{N} \bigr) \, \bop{3} \right) \\
	+ \frac{g_2}{\sqrt{\mathcal{N}} \,} \bigl( \aopd{2} + \aop{2} \bigr) \, \bigl( \bopd{3} \, \bop{2} + \bopd{2} \, \bop{3} \bigr).
\end{multline}
The first line is the free part of the Hamiltonian, from which one can infer the meaning of the HP bosons: The number of HP bosons in the mode with frequency $\delta$ is given by the operator $\bopd{2} \, \bop{2}$. This means that $\bopd{2}$ is related to the creation of excitations with energy $\delta$, which is the energy separation of the single-particle energy levels $\ket{1}$ and $\ket{2}$. Thus, the operator $\bopd{2}$ can be understood as collectively exciting the particles from the first energy level to the second one. This is visualized in Fig.~\ref{fig:HPBosonen}. An analogous reasoning can be given for the other HP boson corresponding to the operator $\bop{3}$.

\subsection{Thermodynamic limit}
The expectation value of the HP boson operators $\bop{r}$ is zero for a finite number $\mathcal{N}$ of particles. In contrast, in the thermodynamic limit the expectation value of this operator can be finite, and is then macroscopic. Given that the occupations $\braket{\Ann}$, and $\braket{\aopd{n} \, \aop{n}}$ should scale with the particle number $\mathcal{N}$, we make the ansatz
\begin{subequations}
	\label{eq:AnsatzSR}
\begin{gather}
	\bop{r} = \sqrt{\mathcal{N} \,} \, \Psi_r + \dop{r}, \quad r \neq m, \\
	\aop{n} = \sqrt{\mathcal{N} \,} \, \varphi_n + \cop{n}, \quad n = 1,2,
\end{gather}
\end{subequations}
in the thermodynamic limit. Here $\sqrt{\mathcal{N} \,} \, \Psi_r$ and $\sqrt{\mathcal{N} \,} \, \varphi_n$ are the ground-state expectation values of $\bop{r}$ and $\aop{n}$, respectively. This means that the ground-state expectation value of the bosonic operators $\dop{r}$ and $\cop{n}$ is zero and, consequently, these operators can be interpreted as quantum fluctuations. Furthermore, they fulfill canonical commutator relations and their matrix elements are of the order of $\mathcal{N}^0$. The parameters $\Psi_r$ and $\varphi_n$ can be chosen real and range from zero to one, which ensures $\braket{\bopd{r} \, \bop{r}} \leq \mathcal{N}$. Another viewpoint is, that the operators $\dop{r}$ and $\cop{n}$ can be generated from $\bop{r}$ and $\aop{n}$ respectively by a canonical transformation and can be considered as displaced bosonic modes \cite{Emary2003}.

Using the ansatz~\eqref{eq:AnsatzSR} we find that the ground-state occupations of the particles and of the scalar bosonic modes are given by

\begin{align}
	\braket{\Axy{r}{r}} &= \mathcal{N} \, \Psi_r^2 + \braket{\dopd{r} \, \dop{r}}, \qquad r \neq m,	\label{eq:OccPartMFa} \\
	\braket{\Axy{m}{m}} &= \mathcal{N} \, \psi_m^2 - \sum_{r \neq m} \braket{\dopd{r} \, \dop{r}},	\label{eq:OccPartMFb} \\
	\braket{\aopd{n} \, \aop{n}} &= \mathcal{N} \, \varphi_n^2 + \braket{\copd{n} \, \cop{n}}, \qquad n = 1,2,	\label{eq:OccBosMF}
\end{align}
with the abbreviation

\begin{equation}
	\psi_m^2 = 1 - \sum_{r \neq m} \Psi_r^2.
\end{equation}

Inserting the ansatz~\eqref{eq:AnsatzSR} into the operator $\hat{\Theta}_m \bigl( \mathcal{N} \bigr)$~\eqref{eq:DefTheta} of the  Holstein--Primakoff transformation~\eqref{eq:DefHP} we obtain

\begin{equation}
	\hat{\Theta}_m \bigl( \mathcal{N} \bigr) = \sqrt{ \mathcal{N} \, \psi_m^2 -\sum_{r \neq m} \bigl[ \dopd{r} \, \dop{r} + \sqrt{\mathcal{N} \,} \, \Psi_r \, \bigl( \dopd{r} + \dop{r} \bigr) \bigr] \; }.
\end{equation}
Since we are working in the thermodynamic limit, we can asymptotically expand the square root in powers of $\sqrt{1/\mathcal{N} \;}$ and obtain up to the order $\mathcal{N}^{-1}$:
\begin{multline}	\label{eq:ThetaThermLim}
	\hat{\Theta}_m \bigl( \mathcal{N} \bigr) \approx \sqrt{\mathcal{N} \,} \, \psi_m \biggl\{ 1 - \frac{1}{2 \sqrt{\mathcal{N} \,} \, \psi_m^2} \sum_{r \neq m} \Psi_r \bigl( \dopd{r} + \dop{r} \bigr) \\
	- \frac{1}{2 \, \mathcal{N} \, \psi_m^2} \sum_{r \neq m} \Bigl[ \dopd{r} \, \dop{r} + \sum_{s \neq m} \frac{\Psi_r \, \Psi_s}{4 \, \psi_m^2} \bigl( \dopd{r} + \dop{r} \bigr) \bigl( \dopd{s} + \dop{s} \bigr) \Bigr] \biggr\}.
\end{multline}
In this expansion we have neglected terms of the order $\mathcal{N}^{-3/2}$ and higher, which do not contribute to the Hamiltonian~\eqref{eq:HamOpHP} in the thermodynamic limit.

Finally, we insert the expression~\eqref{eq:ThetaThermLim} for the operator $\hat{\Theta}_m \bigl( \mathcal{N} \bigr)$ and the ansatz~\eqref{eq:AnsatzSR} into the Hamiltonian~\eqref{eq:HamOpHP}. In the thermodynamic limit we can neglect terms with inverse powers of $\mathcal{N}$ and constants of the order $\mathcal{N}^0$. This eventually yields
\begin{equation}	\label{eq:HamOpHPTL}
	\hat{H}_{m=1} = \mathcal{N} \hat{h}_{m=1}^{(0)} + \sqrt{\mathcal{N} \,} \hat{h}_{m=1}^{(1)} + \hat{h}_{m=1}^{(2)},
\end{equation}
with
\begin{align}
	\hat{h}_{m=1}^{(0)} &=  E_1 + \delta \, \Psi_2^2 + \Delta \, \Psi_3^2 + \omega_1 \, \varphi_1^2 + \omega_2 \, \varphi_2^2 \label{eq:HamOpHPTL0} \\
	& + 4 \, g_1 \, \varphi_1 \, \psi_1 \, \Psi_3  + 4 \, g_2 \, \varphi_2 \, \Psi_2 \, \Psi_3, \nonumber \\
	\hat{h}_{m=1}^{(1)} &= \bigl( \dopd{2} + \dop{2} \bigr) \bigl[ \delta \, \Psi_2 - 2 \, g_1 \, \varphi_1 \, \Psi_2 \, \Psi_3 / \psi_1 \label{eq:HamOpHPTL1} \\
	& + 2 \, g_2 \, \varphi_2 \, \Psi_3 \bigr] + \bigl( \dopd{3} + \dop{3} \bigr) \bigl[ \Delta \, \Psi_3 \nonumber \\
	& + 2 \, g_1 \, \varphi_1 \, \psi_1 \bigl( 1 - \Psi_3^2 / \psi_1^2 \bigr) + 2 \, g_2 \, \varphi_2 \, \Psi_2 \bigr] \nonumber \\
	& + \bigl(\copd{1} + \cop{1} \bigr) \bigl( \omega_1 \, \varphi_1 + 2 \, g_1 \, \psi_1 \, \Psi_3 \bigr) \bigr] \nonumber \\
	& + \bigl(\copd{2} + \cop{2} \bigr) \bigl( \omega_2 \, \varphi_2 + 2 \, g_2 \, \Psi_2 \, \Psi_3 \bigr) \bigr], \nonumber \\
	\hat{h}_{m=1}^{(2)} &= \dopd{2} \, \dop{2} \bigl[ \delta - 2 \, g_1 \, \varphi_1 \, \Psi_3 / \psi_1 \bigr] \label{eq:HamOpHPTL2} \\
	& + \dopd{3} \, \dop{3} \bigl[ \Delta - 2 \, g_1 \, \varphi_1 \, \Psi_3 / \psi_1 \bigr] \nonumber \\
	& + \omega_1 \, \copd{1} \, \cop{1} + \omega_2 \, \copd{2} \, \cop{2} \nonumber \\
	& - \bigl( \dopd{2} + \dop{2} \bigr)^2 \frac{1}{2} g_1 \, \varphi_1 \, \Psi_2^2 \, \Psi_3 / \psi_1^3 \nonumber \\
	& - \bigl( \dopd{3} + \dop{3} \bigr)^2 g_1 \, \varphi_1 \, \Psi_3 / \psi_1 \bigl( 1 + \tfrac{1}{2} \Psi_3^2 / \psi_1^2 \bigr) \nonumber \\
	& - \bigl( \dopd{2} + \dop{2} \bigr) \bigl( \dopd{3} + \dop{3} \bigr) g_1 \, \varphi_1 \, \Psi_2 / \psi_1 \bigl( 1 + \Psi_3^2 / \psi_1^2 \bigr) \nonumber \\
	& + \bigl( \dopd{3} \, \dop{2} + \dopd{2} \, \dop{3} \bigr) 2 \, g_2 \, \varphi_2 \nonumber \\
	& - \bigl( \copd{1} + \cop{1} \bigr) \bigl( \dopd{2} + \dop{2} \bigr) g_1 \, \Psi_2 \, \Psi_3 / \psi_1 \nonumber \\
	& + \bigl( \copd{1} + \cop{1} \bigr) \bigl( \dopd{3} + \dop{3} \bigr) g_1 \, \psi_1 \bigl( 1 - \Psi_3^2 / \psi_1^2 \bigr) \nonumber \\
	& + \bigl( \copd{2} + \cop{2} \bigr) \bigl( \dopd{2} + \dop{2} \bigr) g_2 \, \Psi_3 \nonumber \\
	& + \bigl( \copd{2} + \cop{2} \bigr) \bigl( \dopd{3} + \dop{3} \bigr) g_2 \, \Psi_2. \nonumber
\end{align}
The Hamiltonian $\hat{H}$ separates into three parts $\hat{h}^{(n)}$, each one scaling with $\mathcal{N}^{(2-n)/2}$ and containing products of $n$ operators $\dop{r}$, $\cop{i}$.

\subsection{Ground-state properties} The ground-state energy $\hat{h}_{m}^{(0)}$~\eqref{eq:HamOpHPTL0} of the Hamiltonian~\eqref{eq:HamOpHPTL} is a function of the parameters $\varphi_1$, $\varphi_2$, $\Psi_2$ and $\Psi_3$. Next, we extremize the ground-state energy with respect to theses parameters, i.e. we stipulate
\begin{subequations}
	\label{eq:Extremisation}
\begin{gather}
	\frac{\partial \hat{h}_{m=1}^{(0)}}{\partial \varphi_n} \overset{!}{=} 0, \quad n = 1,2, \label{eq:Extremisationa} \\
	\frac{\partial \hat{h}_{m=1}^{(0)}}{\partial \Psi_r} \overset{!}{=} 0, \quad r = 2,3. \label{eq:Extremisationb}
\end{gather}
\end{subequations}
In the case of $\psi_1$ being finite, this stipulation is equivalent to set the coefficients of the \textsl{linear} Hamiltonian $\hat{h}_{m=1}^{(1)}$~\eqref{eq:HamOpHPTL1} equal to zero (cf. Ref. \cite{Emary2003}).

The first set of Eqs.~\eqref{eq:Extremisationa} gives conditional equations for the parameters $\varphi_n$ of the scalar bosonic modes,
\begin{equation}	\label{eq:phiSol}
	 \varphi_1 = - 2 \frac{g_1}{\omega_1} \, \psi_1 \, \Psi_3, \quad \varphi_2 = - 2 \frac{g_2}{\omega_2} \, \Psi_2 \, \Psi_3,
\end{equation}
which, when inserted into the second set of Eqs.~\eqref{eq:Extremisationb}, gives conditional equations for the parameters $\Psi_r$ of the HP-bosons,
\begin{subequations}
	\label{eq:PsiSol}
\begin{gather}
	\Bigl[ \delta + 4 \Bigl( \frac{g_1^2}{\omega_1} - \frac{g_2^2}{\omega_2} \Bigl) \Psi_3^2 \Big] \, \Psi_2 = 0,	\label{eq:PsiSola} \\
	\Bigl[ \Delta - 4 \frac{g_1^2}{\omega_1} \Bigl( 1 - \Psi_2^2 - 2 \Psi_3^2 \Bigr) - 4 \frac{g_2^2}{\omega_2} \Psi_2^2 \Bigr] \, \Psi_3 = 0.
\end{gather}
\end{subequations}
These equations have several sets of solutions:

(\textit{i}) \textit{Normal state.} The trivial solution, $\Psi_2 = \Psi_3 = 0$, is attended by $\varphi_1 = \varphi_2 = 0$ (cf. Eq.~\eqref{eq:phiSol}). Since $\varphi_n^2$ measures the macroscopic ($\sim \mathcal{N}$) ground-state expectation value of the $n$th scalar bosonic mode (cf. Eq.~\eqref{eq:OccBosMF}), this trivial solution describes the \textit{normal state}, i.e. no superradiant state of the system. In addition, the ground-state expectation value of the occupation of the $n$th energy level, which is given by $\braket{\Ann}$, is macroscopic for $n=1$ only (cf. Eqs.~\eqref{eq:OccPartMFa}, \eqref{eq:OccPartMFb}). Thus, all particles occupy their respective ground state $\ket{1}$. The ground-state energy of the many-particle system is given by $\hat{h}_{\text{normal}}^{(0)} = E_1$. Finally, we note that the normal state is always a solution of the equations~\eqref{eq:phiSol} and \eqref{eq:PsiSol}, irrespective of the couplings $g_1$ and $g_2$. However, analyzing the Hessian matrix of $\hat{h}_{m=1}^{(0)}$ restricts the range of the first coupling to
\begin{equation}
	g_1 < \sqrt{ \Delta \, \omega_1 \,} / 2 \equiv g_{1,c}.
\end{equation}

(\textit{ii}) \textit{Blue superradiant state.} The second solution of Eq.~\eqref{eq:PsiSol} is given by
\begin{subequations}
	\label{eq:SR1Sol}
\begin{align}
	\Psi_2 &= 0, \qquad \Psi_3 = \pm \sqrt{\frac{1}{2} \,} \, \sqrt{ 1 - \Bigl( \frac{g_{1,c}}{g_1} \Bigr)^2 \,}, \\
	\varphi_1 &= \mp \frac{g_1}{\omega_1} \, \sqrt{1 - \Bigl( \frac{g_{1,c}}{g_1} \Bigr)^4 \,}, \qquad \varphi_2 = 0.
\end{align}
\end{subequations}
In contrast to the previous solution, this solution has a finite parameter $\varphi_1$ and for this reason a finite and macroscopic occupation $\braket{\aopd{1} \, \aop{1}}$ of the first scalar bosonic mode. This solution corresponds to a \textit{superradiant} state of the system, where superradiance occurs in the blue branch of the $\Lambda$-system. More precisely, we call this state a \textit{blue superradiant state}. Furthermore, the first \textsl{and} the third single-particle energy level are macroscopically occupied. If we insert the solution~\eqref{eq:SR1Sol} into the ground-state energy~\eqref{eq:HamOpHPTL0} of the many-particle system, we obtain

\begin{equation}	\label{eq:E0blue}
	\hat{h}_{\text{blue}}^{(0)} = E_1 - \frac{\Delta}{4} \Bigl( \frac{g_1}{g_{1,c}} \Bigr)^2 \Bigl[ 1 - \Bigl( \frac{g_{1,c}}{g_1} \Bigr)^2 \Bigr]^2.
\end{equation}
Hence, the ground-state energy of the superradiant state is always smaller than the ground-state energy of the normal state. However, this solution is only valid for couplings $g_1 \ge g_{1,c}$, since for smaller couplings $g_1$ the non-zero parameters of the solution~\eqref{eq:SR1Sol} become purely imaginary and, in addition, the Hessian matrix of $\hat{h}_{m=1}^{(0)}$ becomes indefinite.

(\textit{iii}) \textit{Red superradiant state.} There can be another set of parameters $\varphi_n$, $\Psi_r$ which extremize the ground-state energy $\hat{h}_m^{(0)}$. This set cannot be deduced from the ground-state energy $\hat{h}_{m=1}^{(0)}$ from Eq.~\eqref{eq:HamOpHPTL0}, because it represents not a \textsl{local} but a \textsl{global} minimum of $\hat{h}_{m=1}^{(0)}$. Since $\hat{h}_{m=1}^{(0)}$ is defined on the unit ball $B_2 = \bigl\{ (x,y) \in \mathbb{R}^2 \, \bigr| \, x^2 + y^2 \le 1 \bigr\}$, the global minimum lies on the boundary of $B_2$, that is $\Psi_2^2 + \Psi_3^2 = 1$ ($\psi_1 = 0$) holds. To obtain this global minimum one has to first of all set $\psi_1 = 0$ in Eq.~\eqref{eq:HamOpHPTL0} and omit all terms involving $\psi_1$ in Eq.~\eqref{eq:HamOpHPTL1}. Secondly, one extremizes the ground-state energy as before, but taking the constraint $\Psi_2^2 + \Psi_3^2 = 1$ into account. Eventually, we obtain
\begin{subequations}
\begin{align}
	\Psi_2 &= \pm \sqrt{\frac{1}{2}} \sqrt{ 1 + \Bigl( \frac{g_{2,c_1}}{g_2} \Bigr)^2 }, \\ \Psi_3 &= \pm \sqrt{\frac{1}{2}} \sqrt{ 1 - \Bigl( \frac{g_{2,c_1}}{g_2} \Bigr)^2 }, \\
	\varphi_1 &= 0, \quad \varphi_2 = \mp \frac{g_2}{\omega_2} \sqrt{ 1 - \Bigl( \frac{g_{2,c_1}}{g_2} \Bigr)^4 },
\end{align}
\end{subequations}
where we have introduced
\begin{equation}
	g_{2,c_1} \equiv \frac{\sqrt{ ( \Delta - \delta ) \, \omega_2 }}{2}.
\end{equation}
The occupation of the first energy level $\ket{1}$ is not macroscopic, i.e. it is negligible in the thermodynamic limit. Since $\varphi_1 = 0$ and $\varphi_2$ is finite, this state also corresponds to a superradiant state, whereat superradiance occurs in the red branch of the $\Lambda$-system. We call this superradiant state a \textit{red superradiant} state.

This solution can also be found by direct extremization of the ground-state energy $\hat{h}_{m=2}^{(0)} ( \Psi_1, \Psi_3 )$, i.e. if one considers the second level $\ket{2}$ as the reference state $m$ of the Holstein--Primakoff transformation~\eqref{eq:DefHP}. In general, one can say that using the Holstein--Primakoff transformation in the thermodynamic limit with the $m$th state as the reference state, one can describe many-particle states in which the occupation of the $m$th energy level of the single-particle system is finite. In order to describe the normal state, which is a state where \textsl{all} particles occupy their respective ground state $\ket{1}$, one has to take $\ket{1}$ as the reference state ($m=1$). In contrast, to describe a state where no particle occupies its respective ground state $\ket{1}$, either $\ket{2}$ ($m=2$) or $\ket{3}$ $(m=3)$ has to be chosen as the reference state. Furthermore, we note that one can easily obtain $\hat{h}_{m=2}^{(0)} ( \Psi_1, \Psi_3 )$ from $\hat{h}_{m=1}^{(0)} ( \Psi_2, \Psi_3 )$ from Eq.~\eqref{eq:HamOpHPTL0} by substituting $\psi_1 \mapsto \Psi_1$ and $\Psi_2 \mapsto \psi_2 = \sqrt{ 1 - \Psi_1^2 - \Psi_3^2 \, }$.

At last, the ground-state energy of this red superradiant state is given by

\begin{multline}	\label{eq:E0red}
	\hat{h}_{\text{red}}^{(0)} = E_1 + \delta \\
	- \frac{1}{4} \Biggl[ \Bigl( \sqrt{\Delta} + \sqrt{\delta} \Bigr) \, \frac{g_2}{g_{2,c_2}} - \Bigl( \sqrt{\Delta} - \sqrt{\delta} \Bigr) \, \frac{g_{2, c_2}}{g_2} \Biggr]^2,
\end{multline}
where
\begin{equation}
	g_{2,c_2} \equiv \frac{ \bigl( \sqrt{\Delta} + \sqrt{\delta} \bigr) \sqrt{\omega_2} }{2}
\end{equation}
is a second critical coupling strength.

\textit{Unphysical solution.}
There is also a solution of the Eqs.~\eqref{eq:PsiSol} which corresponds to a state where both branches of the $\Lambda$-system are superradiant. However, this state is either not well defined for certain couplings $g_1$ and $g_2$ or it does \textsl{not} minimize the ground-state energy~\eqref{eq:HamOpHPTL0}. In the latter case, this solution can be attributed to a point of inflection on the energy landscape $\hat{h}^{(0)}_{m=1} ( \Psi_2, \Psi_3 )$.

A further solution of the Eqs.~\eqref{eq:PsiSol} represents a dark state. This state is discussed in detail in Sec.~\ref{sec:DarkState}.

\subsection{Excitation energies}
So far, we have extremize the ground-state energy $\hat{h}^{(0)}$ of the Hamiltonian~\eqref{eq:HamOpHPTL} in the thermodynamic limit. By this procedure the linear part $\hat{h}^{(1)}$ is eliminated as well. The next step is to diagonalize the quadratic part $\hat{h}^{(2)}$. This can be achieved by means of a principle axis or \textit{Bogoliubov} transformation~\cite{Emary2003}. The diagonalized Hamiltonian is then given by

\begin{equation}
	\hat{h}^{(2)} = \sum_k \varepsilon_k \, \eopd{k} \, \eop{k},
\end{equation}
where $\eopd{k} (\eop{k})$ create (annihilate) quasi-particles which refer to bosonic excitations, i.e. $\eopd{k}$ and $\eop{k}$ satisfy canonical commutator relations. The operators $\eopd{k}, \eop{k}$ and the excitation energies $\varepsilon_{k}$ have to be evaluated separately in the three different states. The determination of these quantities reduces to a diagonalization of two-by-two matrices. The diagonalization procedure yields four excitation energies, given by $\bigl( k = ( x \in \{1, 2\}, \sigma \in \{+, -\} ) \bigr)$
\begin{align}
	\varepsilon_{x, \pm}^2 &= \frac{1}{2} \Biggl[ \omega_1^2 + \omega_{1, -}^2 + 2 \lambda \, \omega_{1,-} \label{eq:ExEnergies1} \\
	&\quad \pm \sqrt{ \Bigl( \omega_1^2 - \omega_{1, -}^2 - 2 \lambda \, \omega_{1,-} \Bigr)^2 + 16 \, \tilde{g}_1^2 \, \omega_1 \, \omega_{1,-} \, } \hphantom{i} \Biggr], \nonumber \\
	\varepsilon_{x', \pm}^2 &= \frac{1}{2} \Biggl[ \omega_2^2 + \omega_{2, -}^2 \label{eq:ExEnergies2} \\
	&\qquad \qquad \; \; \pm \sqrt{ \Bigl( \omega_2^2 - \omega_{2, -}^2 \Bigr)^2 + 16 \, \tilde{g}_2^2 \, \omega_2 \, \omega_{2,-} \, } \hphantom{i} \Biggr], \nonumber
\end{align}
with the abbreviations
\begin{subequations}
\begin{gather}
	\omega_{1,-} = \frac{\bar{\Delta}}{2} \big( 1 + \eta_x \bigr), \quad \omega_{2,-} = \bar{\delta} - \frac{\bar{\Delta}}{2} \bigl( 1 - \eta_x \bigr), \\
	\lambda = -\frac{\bar{\Delta}}{8} \frac{(1 - \eta_x) (1 + 3 \, \eta_x)}{1 + \eta_x}, \\
	\tilde{g}_x = g_x \sqrt{ \frac{2}{\eta_x (1 + \eta_x)} }, \quad \tilde{g}_{x'} = \pm g_{x'} \sqrt{\frac{\eta_x - 1}{2 \, \eta_x}}.
\end{gather}
\end{subequations}
It holds for the normal state: $x = 1$, $x' = 2$, $\bar{\Delta} = \Delta$, $\bar{\delta} = \delta$ and $\eta_1 = \eta_2 = 1$; for the blue superradiant state: $x = 1, x' = 2$, $\bar{\Delta} = \Delta$, $\bar{\delta} = \delta$ and $\eta_1 = ( g_1 / g_{1,c} )^2$; and for the red superradiant state: $x = 2, x' = 1$, $\bar{\Delta} = \Delta - \delta$, $\bar{\delta} = -\delta$ and $\eta_2 = \bigl( g_2 / g_{2,c_1} \bigr)^2$.

%-----------------------------------------------------

%-----------------------------------------------------

\section{Phase transitions}
\label{sec:Discussion}

\begin{figure}[tb]
	\includegraphics[width=7cm]{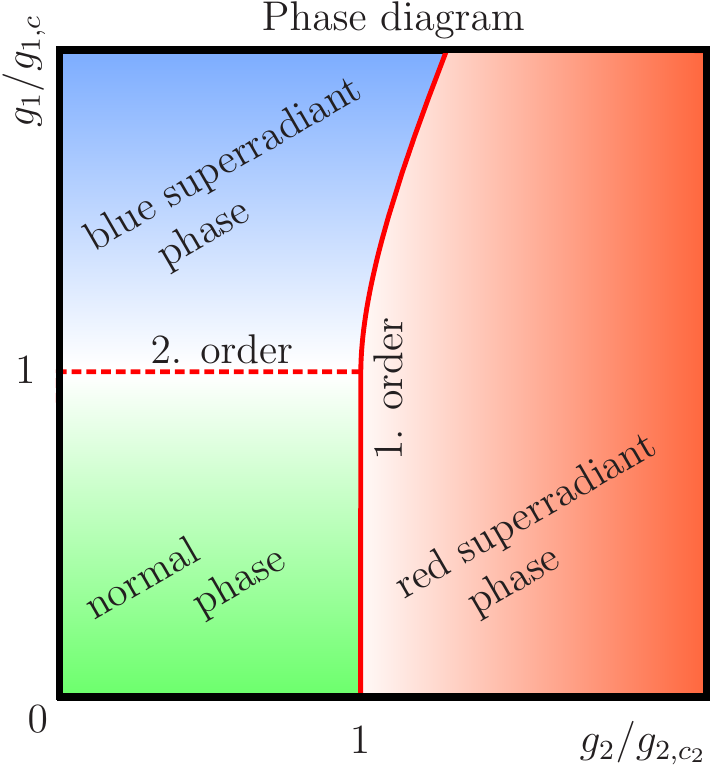}
	\caption{(Color online) Phase diagram for $\Delta > \delta > 0$ showing the three different phases: the normal and the blue and red superradiant phase. The (symmetric) normal phase is defined by $\Psi_2 = \Psi_3 = \varphi_1 = \varphi_2 = 0$. In the (symmetry-broken) blue superradiant phase $\Psi_2 = 0, \Psi_3 \neq 0, \varphi_1 \neq 0$ and $\varphi_2 = 0$. Finally, in the (symmetry-broken) red superradiant phase $\Psi_2 \neq 0, \Psi_3 \neq 0, \varphi_1 = 0$ and $\varphi_2 \neq 0$ holds. The phase transition from the normal to the blue superradiant phase is of second order (red dashed line), whereas the phase transition from the normal to the red superradiant phase and between the two superradiant phases is of first order (red solid line). The normal \textsl{state} is meta stable in the region of the red superradiant phase as long as $g_1 < g_{1,c}$.	\label{fig:phasediagram}}
\end{figure}

Comparing the ground-state energies of the states we found in the last section, we can derive the zero-temperature phase diagram. As mentioned before, the normal state is only stable for couplings $g_1 < g_{1,c}$ and its energy is independent of both coupling strengths $g_1$ and $g_2$. We also observed that the energy of the blue superradiant state is always less than the energy of the normal state. However, the blue superradiant state is stable for $g_1 \ge g_{1,c}$ only. In addition, by comparing the energies of the blue~\eqref{eq:E0blue} and the red~\eqref{eq:E0red} superradiant state, we see that only for $g_2 \ge g_{2,c_2}$ the red superradiant state is stable. Furthermore, in this parameter regime its energy is always smaller than the energy of the normal state (cf. Eq.~\eqref{eq:E0red} with $g_2 = g_{2,c_2}$).

\subsection{Phase diagram}
From this discussion we derive the phase diagram which is shown in Fig.~\ref{fig:phasediagram}. It consists of three phases: one \textit{normal phase} for couplings $g_1 < g_{1,c}$ and $g_2 < g_{2,c_2}$, one \textit{blue superradiant phase} for couplings $g_1 \ge g_{1,c}$ and $g_2 \le \bar{g}_{2,c} (g_1)$, and lastly one \textit{red superradiant phase} for couplings $g_1 < \bar{g}_{1,c} (g_2)$ and $g_2 \ge g_{2,c_2}$. If both couplings are at criticality, $g_1 = g_{1,c}$ and $g_2 = g_{2,c_2}$, all three phases coexist, i.e. there is \textit{triple point} in the phase diagram. Here, $\bar{g}_{1,c} (g_2)$ and $\bar{g}_{2,c} (g_1)$ parameterize the same curve, which represents the phase boundary between the two superradiant phases (see Fig.~\ref{fig:phasediagram}). Both $\bar{g}_{1,c} (g_2)$ and $\bar{g}_{2,c} (g_1)$ are given by the condition that the energies of the blue~\eqref{eq:E0blue} and the red~\eqref{eq:E0red} superradiant state intersect, i.e. both can be obtained by setting the Eqs.~\eqref{eq:E0blue} and \eqref{eq:E0red} equal. For $\bar{g}_{1,c} (g_2)$ we obtain after several algebraic transformations 

\begin{multline}	\label{eq:PhaseBoundary}
	\bar{g}_{1,c}^2 (g_2) = g_2^2 \, \frac{1}{2} \, \frac{\omega_1}{\omega_2} \Biggl\{ 1 + \Bigl( \frac{g_{2,c_1}}{g_2} \Bigr)^4 - \frac{\delta \, \omega_2}{2 \, g_2^2} \\
	+ \Bigl[ 1 + \Bigl( \frac{g_{2,c_1}}{g_2} \Bigr)^2 \Bigr] \sqrt{ \Bigl[ 1 - \Bigl( \frac{g_{2,c_1}}{g_2} \Bigr)^2 \Bigr]^2 - \frac{\delta \, \omega_2}{g_2^2} \phantom{a} } \Biggr\}.
\end{multline}
In the limit $\delta \rightarrow 0$, the phase boundary flattens to a straight line, $\lim_{\delta \rightarrow 0} \bar{g}_{1,c} (g_2) = \sqrt{\omega_1 / \omega_2} \, g_2$.

\begin{figure}[tb]
	\includegraphics[width=8.5cm]{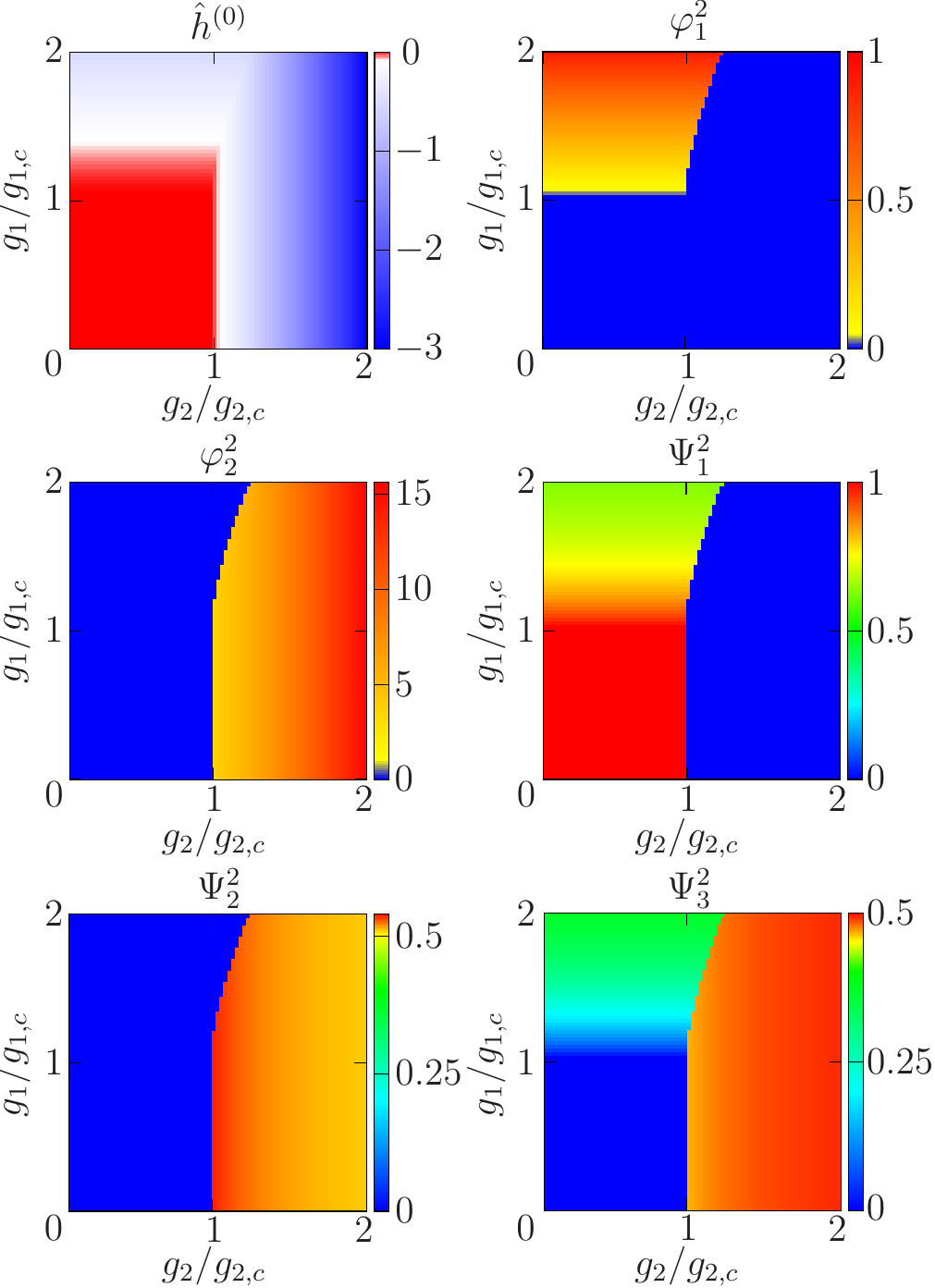}
	\caption{(Color online) Ground state energy $\hat{h}^{(0)}$, ground-state occupation $\varphi_n^2$ of the first ($n=1$) and the second ($n=2$) bosonic mode and the ground-state occupation $\Psi_n^2$ of the single particle energy levels ($n=1,2,3$). Numerical values: $\Delta = \omega_1 = 1$, $\delta = 0.75$, $\omega_2 = 0.25$ (on resonance).	\label{fig:GSOP}}
\end{figure}

The order of a phase transition is defined by the non-analytic behavior of a thermodynamic potential~\cite{Goldenfeld2010}. In the case of zero temperature, the ground-state energy represents a thermodynamic potential and hence its derivatives give the order of the phase transition. The ground-state energy of the normal state is $E_1$ irrespective of the couplings $g_1$ and $g_2$. Hence, all derivatives with respect to $g_1$ and $g_2$ vanish. Comparing this result with the first and second derivatives of the ground-state energy of the blue~\eqref{eq:E0blue} and the red~\eqref{eq:E0red} superradiant state, we see that the phase transition from the normal phase to the blue/red superradiant phase is of second/first order. The ground-state energy is shown in Fig.~\ref{fig:GSOP}.

In addition, the parameters $\Psi_r$ ($r = 2, 3$) and $\varphi_n$ ($n = 1, 2$) also give evidence for the phase transition and can be interpreted as \textit{order parameters}. An order parameter is continuous for second-order phase transitions and discontinuous for first order phase transitions~\cite{Goldenfeld2010}. This behavior is visible in Fig.~\ref{fig:GSOP}. The order parameters are zero in the \textsl{symmetric} (normal) phase and are finite in the \textsl{symmetry-broken} (superradiant) phase. The corresponding symmetry is the parity symmetry (see Sec.~\ref{sec:Model}). In the blue/red superradiant phase the parity symmetry corresponding to the parity operator $\hat{\Pi}_1$/$\hat{\Pi}_2$ (see Eq.~\eqref{eq:DefParity}) is broken, since e.g. in the blue superradiant phase for finite $\varphi_1$ the operator $\copd{1} \, \cop{1}$ in the Hamiltonian~\eqref{eq:HamOpHPTL2} is not invariant under the symmetry transformation $\hat{\Pi}_1$: $\hat{\Pi}_1 \, \copd{1} \, \cop{1} \, \hat{\Pi}_1^\dag = \copd{1} \, \cop{1} + \sqrt{\mathcal{N}} \, \varphi_1 ( \copd{1} + \cop{1} ) + \mathcal{N} \, \varphi_1^2$.

\begin{figure}[tb]
	\includegraphics[width=8.5cm]{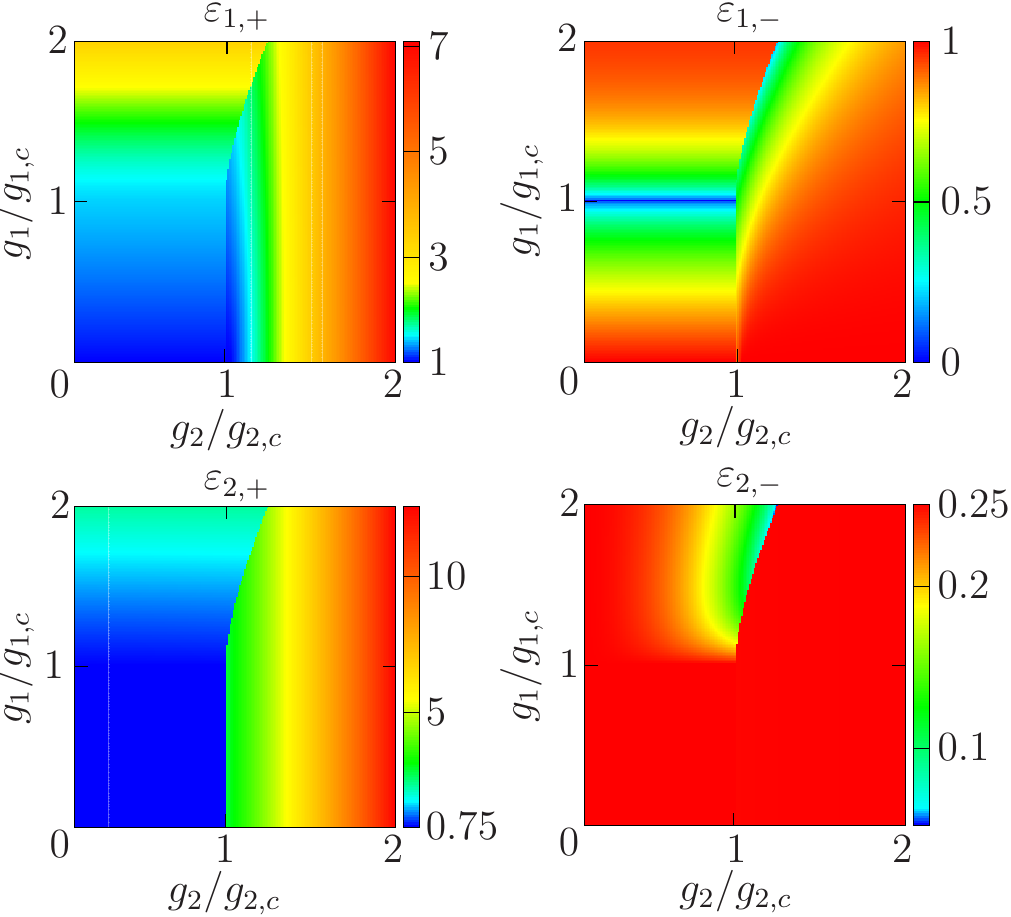}
	\caption{(Color online) Excitation energies $\varepsilon_{n,\sigma}$ from Eqs.~\eqref{eq:ExEnergies1}, \eqref{eq:ExEnergies2}. Numerical values: $\Delta = \omega_1 = 1$, $\delta = 0.75$, $\omega_2 = 0.25$ (on resonance).	\label{fig:ExEnergies}}
\end{figure}

Both, the phase transition and the order of the phase transition can also be deduced from the excitation energies. The excitation energies from the Eqs.~\eqref{eq:ExEnergies1} and~\eqref{eq:ExEnergies2} are shown in Fig.~\ref{fig:ExEnergies}. At the phase transition at least one of the excitation energies either tends to zero or is discontinuous. The first case corresponds to a second-order, the latter case to a first-order phase transition. The second-order phase transition can be read off the excitation energy $\varepsilon_{1,-}$ which is zero for $g_1 = g_{1,c}$ and $g_2 < g_{2,c_2}$.

Finally, we note that the phase transition from the normal to the blue superradiant phase is in accordance with the superradiant phase transition in the Dicke model~\cite{Emary2003}, i.e. it is of second order and one (atomic) branch of the excitation energies tends to zero at the phase transition. The discontinuity of the order parameters and the first derivative of the ground-state energy at the phase transition between the normal and the red superradiant phase scales with $\sqrt{\delta}$. Thus, this first-order phase transition becomes continuous in the limit $\delta \rightarrow 0$. However, the phase boundary between the two superradiant phases persists to be a first-order phase transition in this degenerate limit. This is also the case in the limit of large couplings, $g_1/g_{1,c}, g_2/g_{2,c_2} \rightarrow \infty$.

%-----------------------------------------------------

%-----------------------------------------------------

\subsection{Dark state}
\label{sec:DarkState}

Due to the interaction of a quantum system with its environment decay processes within the quantum system occur. Eigenstates of the Hamiltonian which are unaffected by these decay processes are called dark states. In our model, a dark state is a many-body state which does not radiate, i.e. a state where the occupation in either of the bosonic modes is zero. This condition is satisfied if the two parameters $\varphi_n$ are zero. From the Eqs.~\eqref{eq:phiSol} we see, that the normal state, with $\Psi_2 = \Psi_3 = 0$, is a trivial dark state. In general, it suffices to set $\Psi_3 = 0$ for a dark state. Applying this condition to the Eq.~\eqref{eq:PsiSol}, we can identify a dark state for $\delta = 0$ only, i.e. for two energetically degenerate ground states. In the thermodynamic limit the coherence of this dark state is given by $\braket{\Axy{1}{2}} = \mathcal{N} \, \psi_1 \, \Psi_2$, and is therefore finite apart from the two trivial cases $\psi_1 = 0$ or $\Psi_2 = 0$. The energy of the dark state is simply $\hat{h}_{\text{Dark}}^{(0)} = E_1$.

We obtain the excitation energies for the dark state by diagonalizing $\hat{h}^{(2)}_{m=1}$ from Eq.~\eqref{eq:HamOpHPTL2}. For any given $\Psi_2$, these energies can be computed from the characteristic equation

\begin{equation}
	\det \begin{pmatrix}
			\Delta^2 - \varepsilon^2 & 2 \, g_1 \, \psi_1 \sqrt{\omega_1 \, \Delta \,} & 2 \, g_2 \, \Psi_2 \sqrt{\omega_2 \, \Delta \,} \\
			2 \, g_1 \, \psi_1 \sqrt{\omega_1 \, \Delta \,} & \omega_1^2 - \varepsilon^2 & 0 \\
			2 \, g_2 \, \Psi_2 \sqrt{\omega_2 \, \Delta \,} & 0 & \omega_2^2 - \varepsilon^2
		\end{pmatrix} = 0,
\end{equation}
where $\det M$ is the determinant of the matrix $M$. The characteristic equation is readily solved in the case of two-photon resonance ($\omega_1 = \omega_2 = \Delta \equiv \omega$), yielding the energies

\begin{gather}	\label{eq:DSExciations}
	\varepsilon_0 = 0, \quad \varepsilon_1 = \omega, \\
	\varepsilon_{2, \pm} = \sqrt{ \omega^2 \pm 2 \, \omega \sqrt{g_1^2 \bigl( 1 - \Psi_2^2 \bigr) + g_2^2 \, \Psi_2^2 \,} \,},
\end{gather}
where the additional zeroth mode $\varepsilon_0$ stems from the limit $\delta \rightarrow 0$.

The parameter $\Psi_2$ is arbitrary and can range from zero to one. For a given $\Psi_2$, we find by analysis of the Hessian matrix of $\hat{h}_{m=1}^{(0)}$, that this dark state is meta-stable if the inequality

\begin{equation}	\label{eq:DSStability}
	\bigl( g_1 / g_{1,c} \bigr)^2 \bigl( 1 - \Psi_2^2 \bigr) \leq 1 - \bigl( g_2 / g_{2,c} \bigr)^2 \, \Psi_2^2
\end{equation}
is satisfied. Otherwise this dark state solution is unstable. In Eq.~\eqref{eq:DSStability} we have introduced the critical coupling strength $g_{2,c} \equiv \sqrt{\Delta \, \omega_2 \,} / 2$.

We emphasize that the dark state exists for $\delta = 0$ only. By inspection of the inequality~\eqref{eq:DSStability}, we make the following statements: First, the dark state is stable for $g_1 < g_{1,c}$ \textsl{or} $g_2 < g_{2,c}$ only. Furthermore, if both coupling strengths fulfill $g_n < g_{n,c}$, i.e. in the normal phase, both $\psi_1$ and $\Psi_2$ can range from zero to one. On the other hand, if $g_2 > g_{2,c}$ and $g_1 < g_{1,c}$, then $\Psi_2$ is restricted to the interval $[0,\Psi_{2,\text{max}}]$, where $\Psi_{2,\text{max}} > 0$ is given by the inequality~\eqref{eq:DSStability}. Correspondingly $\psi_1$ is restricted to the interval $[\psi_{1,\text{min}},1]$, with $\psi_{1,\text{min}}$ given by $\sqrt{1 - \Psi_{2,\text{max}}^2 \,}$. An analogue argument can be given for the case $g_1 > g_{1,c}$ and $g_2 < g_{2,c}$, where $\psi_1$ and $\Psi_2$ are interchanged.

For couplings $g_1 \gg g_{1,c}$ and $g_2 < g_{2,c}$, inequality~\eqref{eq:DSStability} restricts the order parameters to $\psi_1 \approx 0$ and $\Psi_2 \approx 1$, i.e. only the second single-particle level is macroscopically occupied. On the other hand, for couplings $g_2 \gg g_{2,c}$ and $g_1 < g_{1,c}$ only the first single-particle level is macroscopically occupied, i.e. $\psi_1 \approx 1$ and $\Psi_2 \approx 0$. This `counterintuitive' behavior is reminiscent of the \mbox{STIRAP} scheme~\cite{Bergmann1998}. In contrast to the \mbox{STIRAP} scheme, the actual values of the populations $\psi_1$ and $\Psi_2$ in this dark state are \textsl{not} defined by the coupling strengths $g_1$ and $g_2$, but rather by the preparation of the system. Thus, the system cannot be driven coherently from a state with all particles occupying the first single-particle energy level $\ket{1}$ to a state where all particles occupy the second single-particle energy level $\ket{2}$ just by changing the couplings.

\begin{figure}[tb]
	\includegraphics[width=4cm]{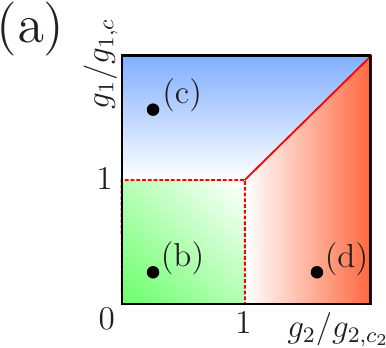} \hfill
	\includegraphics[width=4cm]{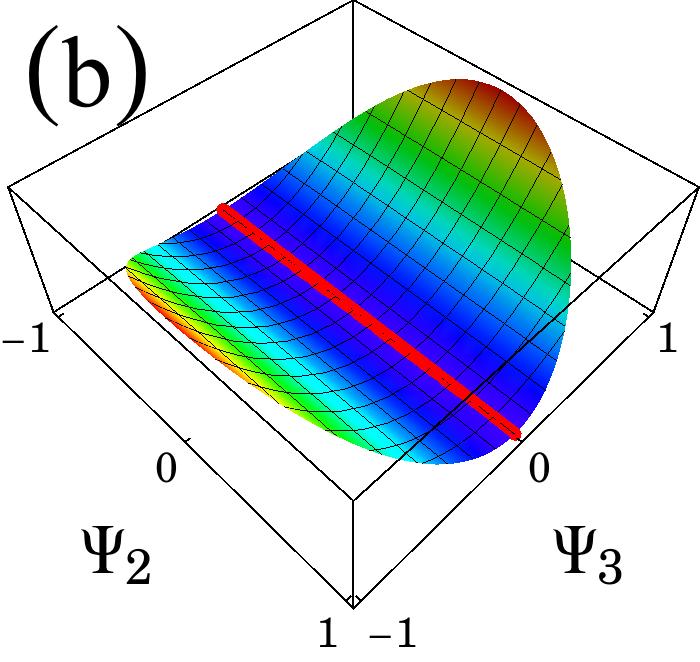} \\ \vspace{0.3cm}
	\includegraphics[width=4cm]{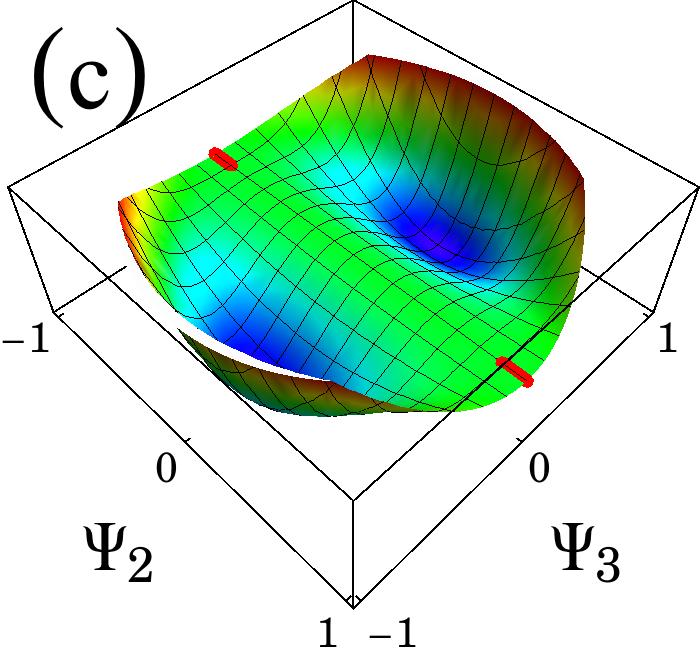} \hfill
	\includegraphics[width=4cm]{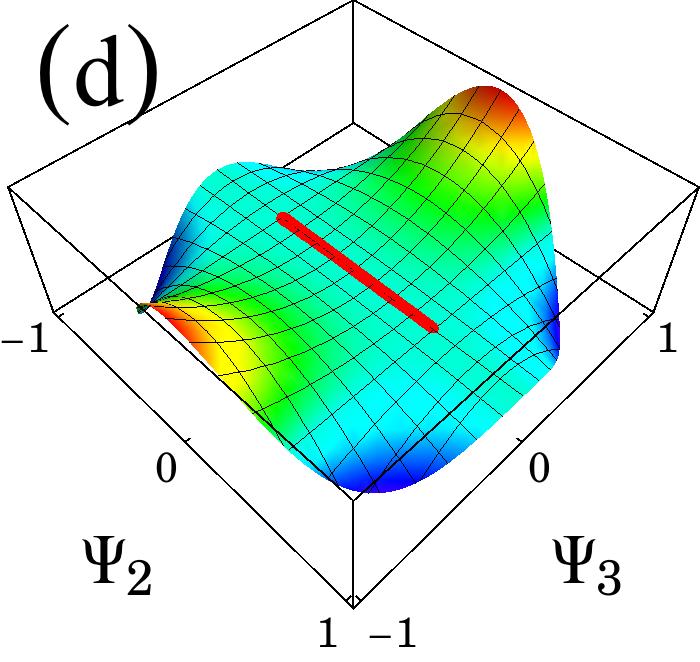}
	\caption{(Color online) Phase diagram (a) as in Fig.~\ref{fig:phasediagram} and energy surfaces $\hat{h}^{(0)}$ (b--d) in the degenerate ($\delta \rightarrow 0$) limit. In (a), the solid and the broken red lines denote first and second-order phase transitions, respectively. The red lines in (b--d) visualize the inequality~\eqref{eq:DSStability}, i.e. the line where the dark state with $\Psi_3 = 0$ is stable. However, as the red line is flat, fluctuations along the $\Psi_2$-direction transform stable states on the red line into unstable states outside the red line. Eventually, these states decay to superradiant states with $\Psi_3 \neq 0$. 	\label{fig:DSEnergySurface}}
\end{figure}

In addition, we note that in this dark state the mode $\varepsilon_0 = 0$ in direction of $\Psi_2$ of the energy surface $\hat{h}^{(0)} ( \Psi_2, \Psi_3 )$ is trivially massless (cf. Eq.~\eqref{eq:DSExciations}). Therefore, tiny fluctuations can easily excite this dark state along the direction of $\Psi_2$, making the state eventually unstable. This instability is visualized in Fig.~\ref{fig:DSEnergySurface}

%-----------------------------------------------------

%-----------------------------------------------------

\section{Conclusions}
\label{sec:Conclusions}

We have analyzed an extension of the well known Dicke model from two to three-level particles. By means of a Holstein--Primakoff transformation we have identified three stable states in the thermodynamic limit: a normal, a blue superradiant, a red superradiant state. At zero temperature these states correspond to three thermodynamic phases, which we have arranged in a phase diagram. The phase transition between the normal and the blue superradiant phase is of second order and all other phase transitions are of first order. We have also shown that a state with both superradiant states coexisting is not stable. A dark state with zero occupancy of the third single-particle level exists for $\delta = 0$ only. However, this dark state is not stable.

As in the original Dicke model, the same experimental difficulties arise in our extended Dicke model, i.e. reaching the critical coupling strength is challenging as well. Hence, using three level atoms has no advantage over the use of two level atoms.

However, we expect that similarly to the Dicke model and its realization in the experiments of \citet{Baumann2010} there should be experimental manageable systems, which can theoretically be described by an effective Hamiltonian of the form presented here. In the case of the experiments in Ref.~\cite{Baumann2010}, this might be achieved by coupling a Bose--Einstein condensate to an additional cavity mode. Furthermore, an even richer phase diagram with additional superradiant phases could be generated in such a system.

Considering a cold quantum gas in an optical lattice, a characteristic feature of our extended Dicke model especially in the degenerate limit, $\delta \rightarrow 0$, could appear. In this regard, we have an extension of a system proposed by \citet{Silver2010} in mind. There, it was shown that a two-band zero-hopping Bose--Hubbard model coupled to a cavity light field can be written as an effective Dicke model. If one superposes a superlattice of twice the wavelength of the original lattice, and couples the superlattice to two independent cavity light fields, this extended Bose--Hubbard model can be mapped to our extended Dicke model with $\delta=0$. Since in experiment one has an extensive control over the parameters of cold quantum gases, the observation of superradiant phases should be feasible.

%-----------------------------------------------------

%-----------------------------------------------------

\section*{Acknowledgments}
We thank Martin Aparicio, Victor Bastidas and Christian Nietner for useful discussions. The work was supported by the Deutsche Forschungsgemeinschaft within the SFB 910 and the project BR 1528/8-1.

%-----------------------------------------------------

\end{document}